\documentclass[a4paper]{jpconf}
\usepackage{graphicx}
\begin{document}
\title{Minimisation strategies for the determination of parton density
  functions}

\author{Stefano Carrazza$^1$ and Nathan
  P. Hartland$^2$}

\address{$^1$ Theoretical Physics Department, CERN, CH-1211 Geneva,
  Switzerland} \address{$^2$ Department of Physics and Astronomy, 
  Vrije Universiteit Amsterdam, NL-1081 HV and Nikhef Theory Group, Science
  Park 105, 1098 XG Amsterdam, The Netherlands}

\ead{stefano.carrazza@cern.ch\footnote{Preprint: CERN-TH-2017-241},n.p.hartland@vu.nl}

\begin{abstract}
  We discuss the current minimisation strategies adopted by research projects
  involving the determination of parton distribution functions (PDFs) and
  fragmentation functions (FFs) through the training of neural networks. We
  present a short overview of a proton PDF determination obtained using the
  covariance matrix adaptation evolution strategy (CMA-ES) optimisation
  algorithm. We perform comparisons between the CMA-ES and the standard
  nodal genetic algorithm (NGA) adopted by the NNPDF collaboration.
\end{abstract}

\section{Introduction}

In perturbative QCD, parton distribution functions are used to describe the
nonperturbative structure of hadrons. These functions are typically determined
by means of a fit to a wide set of experimental data~\cite{Gao:2017yyd}. Such a
fit is complicated by the non-trivial relationship between parton distributions
and physical observables. Data on hadron-hadron collider processes is related to
parton distributions by means of a double convolution with a kernel computed in
perturbation theory. Given some representation of PDFs: $f$ in terms of a set of
model parameters: $\{a\}$ a typical hadron-hadron collider observable $T$ may be
be calculated as

\begin{equation}
    T[f\{a\}] =
    \int{\widetilde{T}_{ij}(x_1,x_2)f_{i/1}(x_1,\{a\})f_{j/2}(x_2,\{a\}),}
        \label{eq:hadronicObs}
\end{equation}
where $f_{i/1}$ and $f_{j/2}$ represent the parton distributions for PDF
flavours $i/j$ of the first and second beams respectively, and $\widetilde{T}$
is the relevant integration kernel for the observable $T$. The model parameters
may then be determined by a fit to data by minimising some figure of merit, for
example the correlated $\chi^2$ between theory and data

\begin{equation}
  \chi^2(\{a\})
  =
  \sum_{i,j}^{N_{\rm dat}}
  (T_i[f]-D_i) (C_{ij})^{-1} (T_j[f]-D_j)
  \,\mbox{,}
\label{eq:chi2def}
\end{equation}
where $D_i$ represents the $i^{\mathrm{th}}$ data point in a set with covariance
$C$. For the most part, in PDF determinations this minimisation procedure is
performed with the use of standard gradient descent methods. However the
computation of the fitness gradient with respect to model parameters, i.e.
\begin{equation} \frac{\partial \chi^2}{\partial a}, \end{equation} is
challenging due to the relationship between theory predictions and PDFs
expressed in Equation~\ref{eq:hadronicObs}.  For fits of PDF models with
relatively few parameters, such gradients may be either directly computed or
reasonably approximated by the equivalent finite-differences. However when using
models with a relatively large parameter space, such as those used in
determinations in the NNPDF approach, the computation or approximation of these
gradients can be a considerable computational burden.  Consequently in NNPDF
determinations of parton distributions, \emph{gradient-free} minimisation
methods have been employed.  These typically take the form of a \emph{Nodal
Genetic Algorithm} (NGA) variant~\cite{Montana:1989}. At each iteration of the
minimisation, candidate parameter values are sampled around a search centre
according to a Gaussian distribution, with the candidate minimising the $\chi^2$
being selected for the search centre in the next iteration. Such an algorithm is
\emph{greedy} in that it always makes the locally optimal selection for the next
iteration. For details of the algorithm see Ref.~\cite{Ball:2014uwa}.

Despite the simplicity of the algorithm, it has proven to be an efficient way of
exploring the complicated PDF parameter space in a manner that is relatively
robust in terms of avoiding features such as local minima. However there are
clear directions for improvement in terms of a minimisation algorithm. Firstly,
the algorithm does not make use of any information on the structure of the
parameter space in order to guide the exploration in subsequent steps. Secondly
the current algorithm is rather sensitive to noise in the objective function, as
it is constrained to only select the locally best candidate at each iteration. 

In the NNPDF approach, the inherent susceptibility of the minimisation
algorithm to noise is ameliorated by two mechanisms.  Each PDF determination is
performed as an ensemble of fits to different sub-samples of the complete
dataset. This ensemble determination reduces the sensitivity of the overall
result to fluctuations affecting individual fits, however there is the potential
for PDF uncertainties to be inflated due to overly noisy ensemble members.
Furthermore a form of cross-validated early stopping is applied to avoid
over-fitting. This does not however completely prevent contamination of the
results by noise.

In a recent application of the NNPDF methodology to the determination of
fragmentation functions~\cite{Bertone:2017tyb} an alternative minimisation
strategy was employed, the active $(\mu,\lambda)$-CMA-ES
algorithm~\cite{hansen2001ecj,Hansen16a}. The CMA-ES procedure includes
mechanisms which overcome the two weaknesses in the standard NNPDF NGA
minimisation described above. In the application to fragmentation functions,
CMA-ES proved a reliable and efficient minimisation algorithm. Here we shall
report results from a preliminary investigation into the application of the
algorithm in the case of proton PDFs.

\section{Comparison of minimisers}

Two fits have been performed at NNLO in the NNPDF3.1
framework~\cite{Ball:2017nwa}, one utilising the NGA minimiser, and the second
with the CMA-ES algorithm with population $\mu=80$ and initial step-size
$\sigma=0.1$. All other CMA-ES parameters are set as per the recommended
defaults. The dataset and other fit parameters are identical to those used in
the NNPDF3.1 NNLO determination. 

\begin{figure} \begin{center}
        \includegraphics[width=0.49\textwidth]{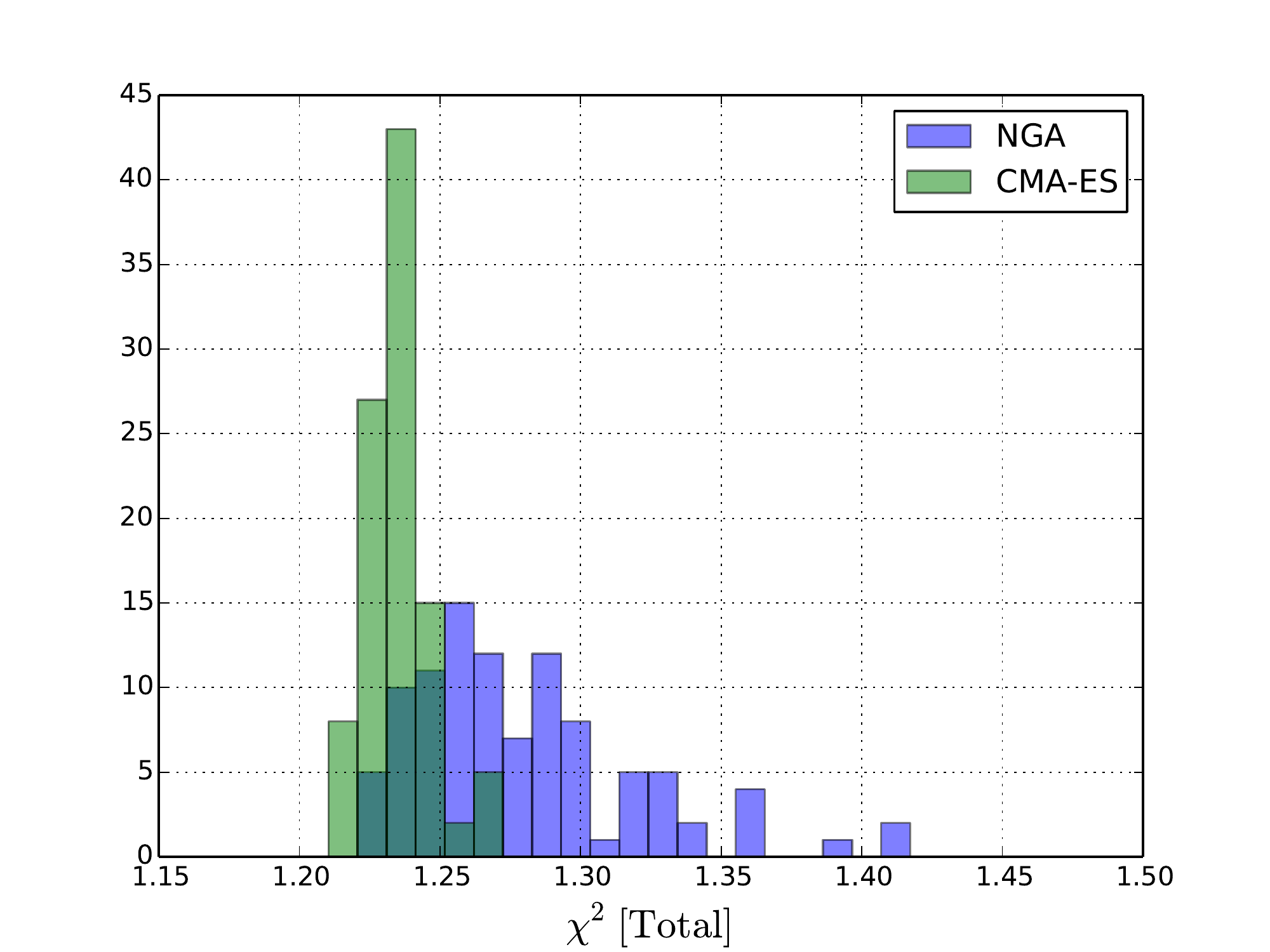}
        \includegraphics[width=0.49\textwidth]{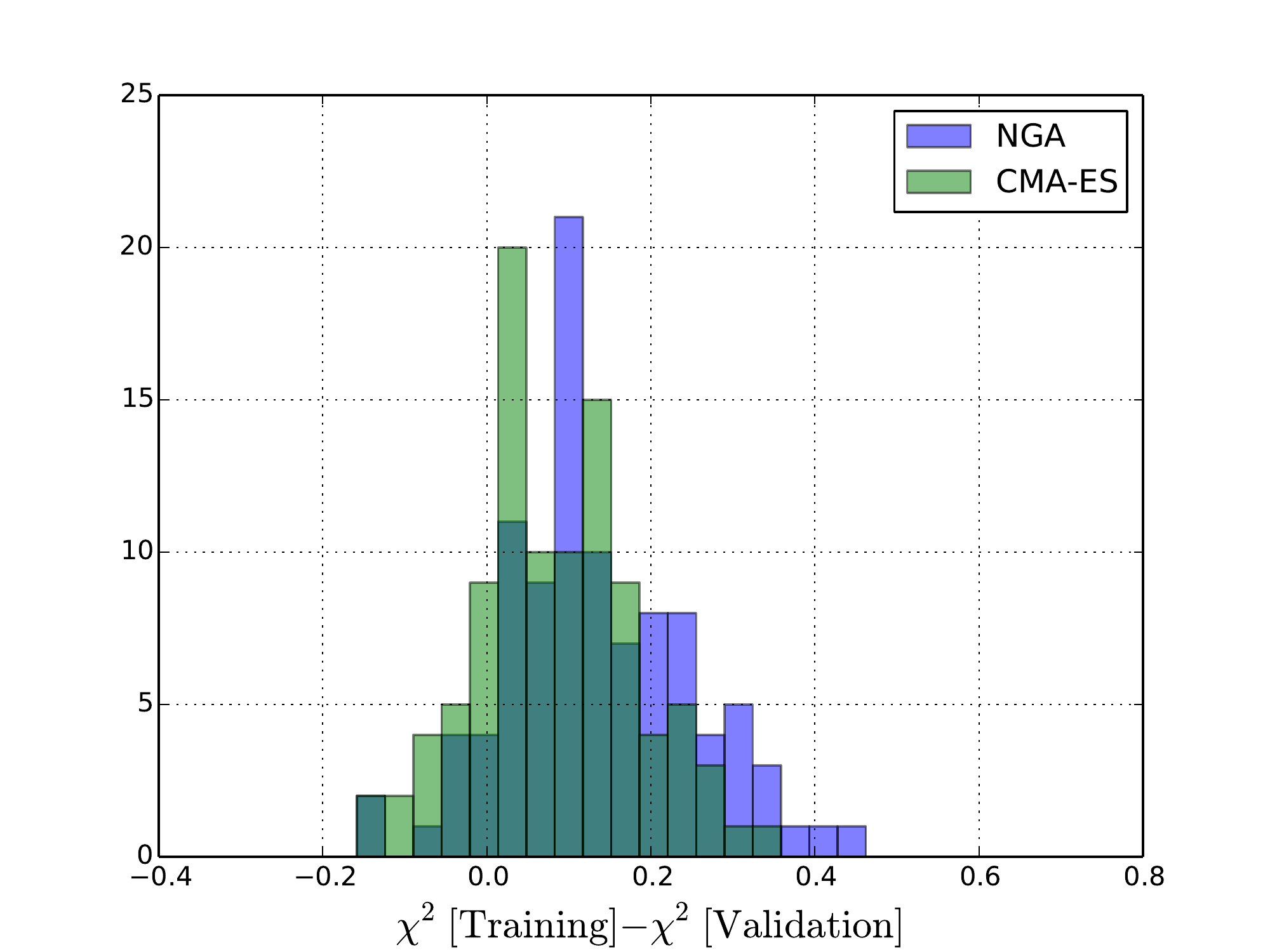}
    \end{center}       \caption{Histogram of $\chi^2$ fitness to the total
        dataset (left) and the difference between training and validation
        $\chi^2$ (right). Shown are the fitnesses of the CMA-ES ensemble (blue)
        and the NNPDF ensemble (green).}\label{fig:chisquared}
\end{figure}

In the first panel of Figure~\ref{fig:chisquared} we show the distribution of
$\chi^2$ fitness over the two PDF ensembles.  The total $\chi^2$ distribution
demonstrates a significant difference between the results from the two
minimisers.  Ensemble members of the CMA-ES fit are typically better fit to
their datasets than the NGA equivalent.  Furthermore the results from the CMA-ES
fit are considerably more consistent between ensemble members. To verify that
this improvement in total fitness is not due to over-learning the training
dataset, in the second panel we histogram the differences between the $\chi^2$
to the datasets used for training and validation. While results here are more
consistent between the two minimisers, it is clear that the CMA-ES results have
a slightly better balance between the training and validation fitnesses than in
the NGA equivalent.

To investigate the greater ensemble consistency demonstrated
in the CMA-ES $\chi^2$ distribution, the arc-length defined as 
\begin{equation}
    L = \int{ \sqrt{1+\left(\frac{df}{dx}\right)^2} dx,}
\end{equation} 
is a useful quantity in that it offers a measure of the structural `complexity'
of a function. For a fit determining an underlying function that is expected to
be largely smooth, results from fits suffering from contamination by noise in
the objective function would be expected to have a typically larger arc-length
than those less susceptible to noise.

\begin{figure}[htbp] 
    \begin{center}
        \includegraphics[width=0.85\textwidth]{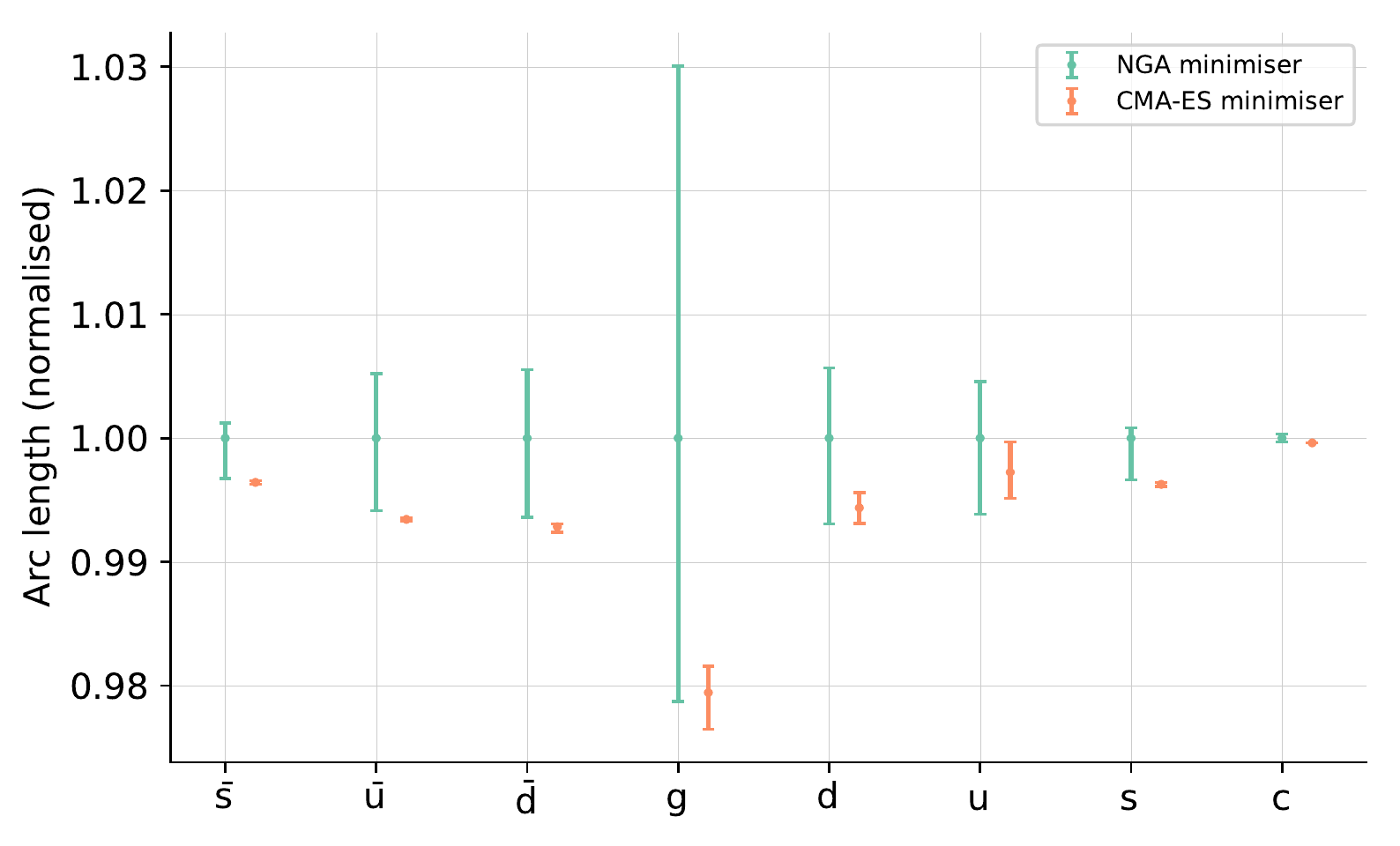}
    \end{center}
    \caption{Arc-lengths $L$ for PDFs determined via the NGA and CMA-ES
        minimisers~. Results are normalised to the NGA value, and uncertainties
        are given as asymmetric 68\% confidence intervals.}\label{fig:arclength} 
\end{figure}

\begin{figure}[htp]
    \begin{center}
        \includegraphics[width=0.49\textwidth]{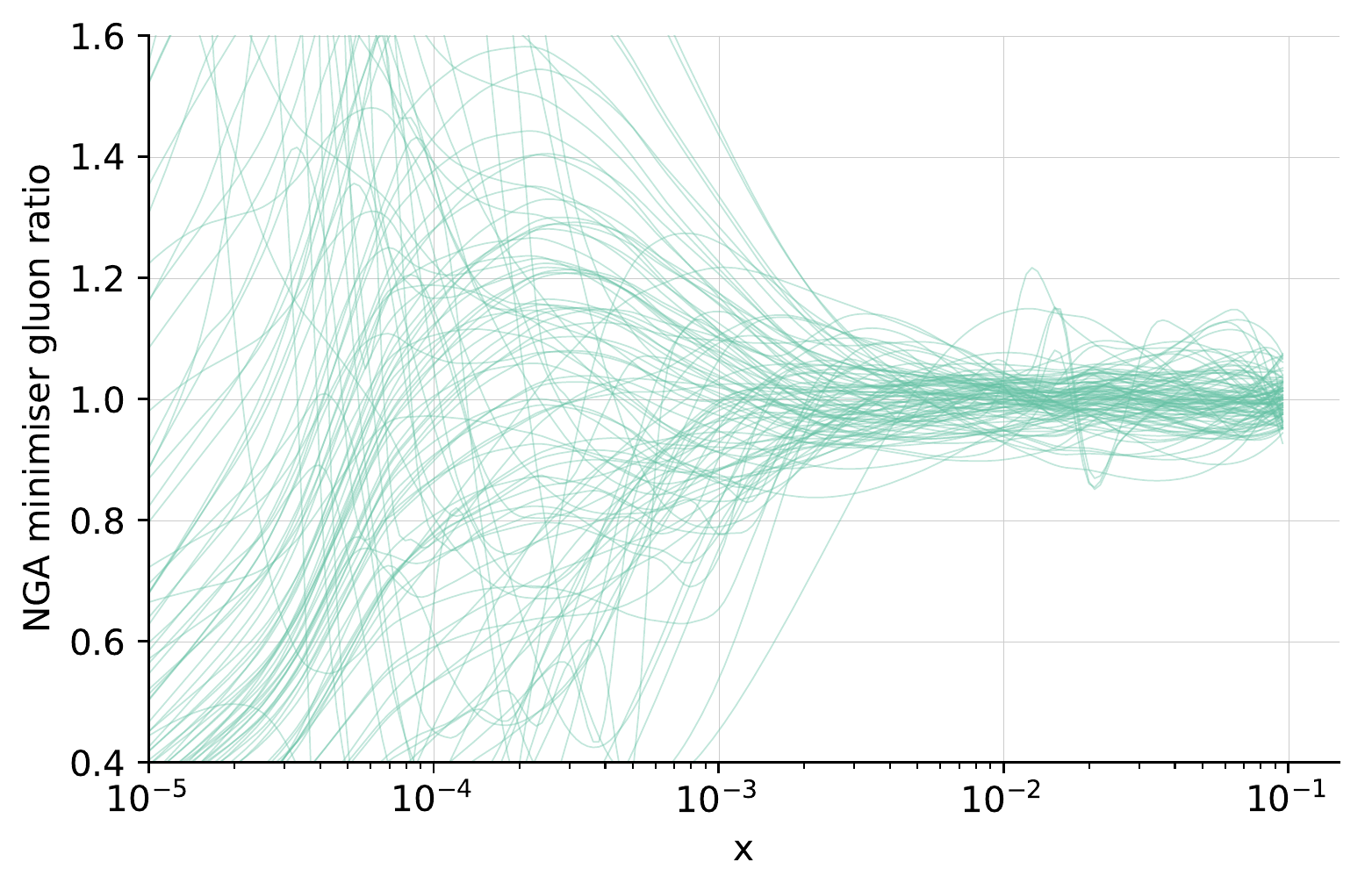}
        \includegraphics[width=0.49\textwidth]{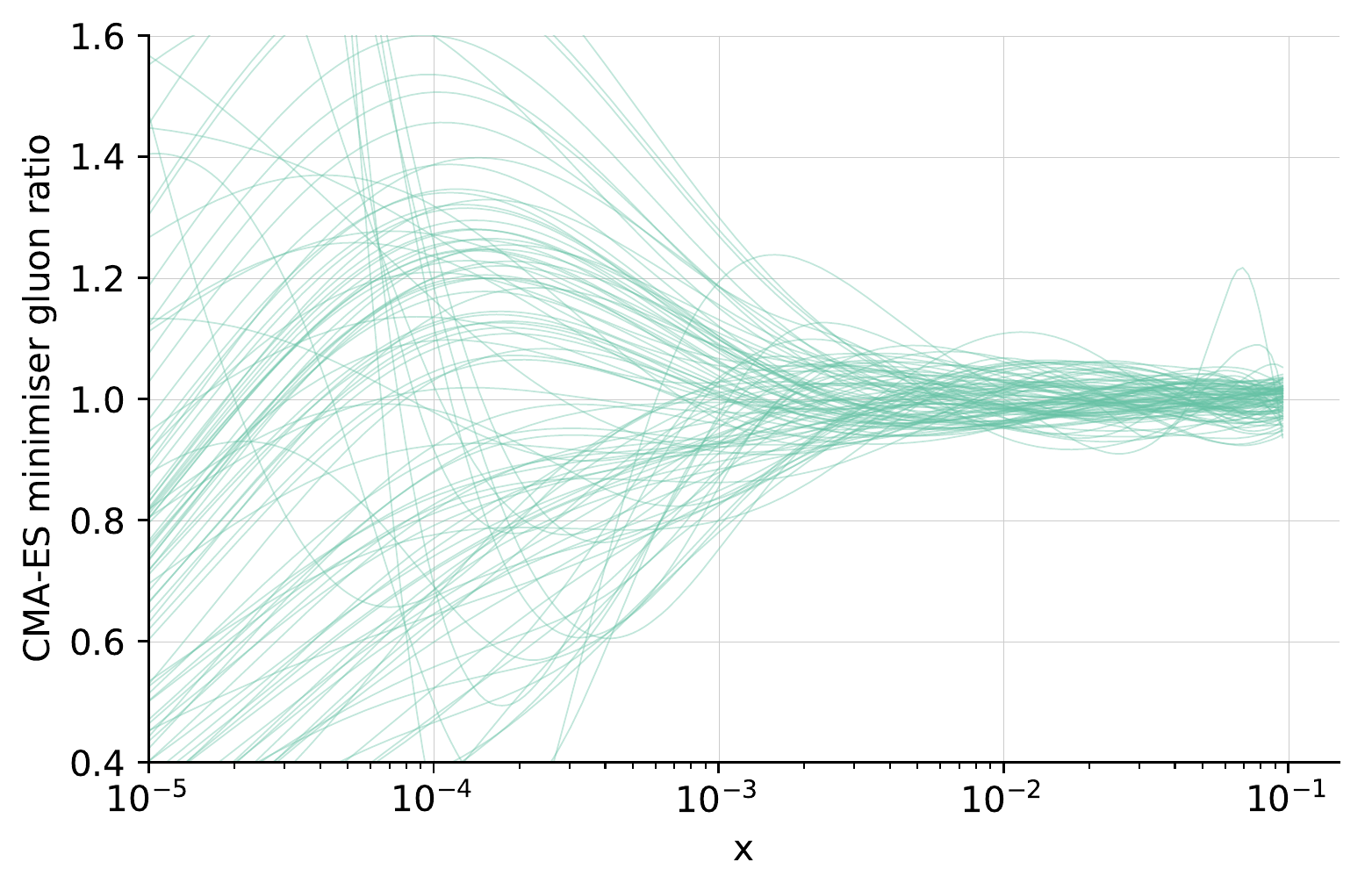}
    \end{center}
    \caption{Gluon PDF ensemble distributions for the NGA
        (left) and the CMA-ES (right) minimisers. Each distribution is
        normalised to its central-value. }\label{fig:pdfs} 
\end{figure}

In Figure~\ref{fig:arclength} we compare the arc-lengths of the PDFs determined
by the NGA minimiser and the CMA-ES~.  The figure demonstrates a clear and
significant difference in both the PDF arc-lengths and their spread over the fit
ensemble. The CMA-ES results demonstrate consistently lower and more regular
arc-lengths, confirming the expectation that the results from the CMA-ES should
be more resistant to noise in the objective function. Particularly striking is
the difference in arc-length for the gluon PDF~. A direct comparison of the gluon
determination in the CMA-ES and NGA minimisers is shown in
Figure~\ref{fig:pdfs}, where even though the results are consistent between the
two minimisers, the reduced structural complexity of the CMA-ES result can be
clearly seen. 


\section{Conclusion} The CMA-ES minimiser has been successfully applied to the
determination of parton distribution functions in the NNPDF approach.  Results
determined through the CMA-ES procedure exhibit greater consistency, improved
agreement with data, and reduced complexity in comparison to results determined
through the NGA minimiser. While these improvements are clear from performing
fits to real data, the statistical consistency of the results are yet to be
established. In terms of arc-lengths, while lower structural complexity may
indicate a reduced sensitivity to noise in the objective function, it may also
indicate an overly restrictive minimisation procedure whereby uncertainties may
be underestimated. In order to ensure the statistical consistency of the
results, further tests are needed. In Ref.~\cite{Ball:2014uwa} the consistency
of the PDF uncertainties were determined by means of statistical closure tests.
For the minimiser investigated here to provide reliable fits of PDFs, it must be
able to demonstrate closure on a pseudo-dataset. These tests we leave for a
future work.

\section*{Acknowledgements}

S.~C. is supported by the HICCUP ERC Consolidator grant (614577) and
by the European Research Council under the European Union's Horizon
2020 research and innovation programme (grant agreement n$^{\circ}$
740006). N.~H. is supported by an European Research Council Starting
Grant ``PDF4BSM''.

\end{document}